\documentclass{article}
\PassOptionsToPackage{numbers,sort& compress}{natbib}
    \usepackage[final]{neurips_2022}

\usepackage[utf8]{inputenc} % allow utf-8 input
\usepackage[T1]{fontenc}    % use 8-bit T1 fonts
\usepackage{hyperref}       % hyperlinks
\usepackage{url}            % simple URL typesetting
\usepackage{booktabs}       % professional-quality tables
\usepackage{amsfonts}       % blackboard math symbols
\usepackage{nicefrac}       % compact symbols for 1/2, etc.
\usepackage{microtype}      % microtypography
\usepackage{xcolor}         % colors
\usepackage{amsmath,amssymb,amsfonts}
\usepackage{algorithmic}
\usepackage{graphicx}
\usepackage{framed,multirow}

\title{Multi-task Learning for Optical Coherence Tomography Angiography (OCTA) Vessel Segmentation}

\author{
\centering
\hspace{-18mm}
\begin{tabular}{ccc}
Can Koz & Onat Dalmaz & Mertay Dayanc \\
Department of Computer Science & Department of Electrical Engineering & Google \\
University of Oxford, OX, UK & Stanford University, CA, USA & Mountain View, CA, USA \\
\texttt{{can.koz}@some.ox.ac.uk} & \texttt{{onat}@stanford.edu} & \texttt{{mertaydayancc}@gmail.com}
\end{tabular}
}
\begin{document}

\maketitle
\vspace{-3mm}

\begin{abstract}
\vspace{-1mm}
Optical Coherence Tomography Angiography (OCTA) is a non-invasive imaging technique that provides high-resolution cross-sectional images of the retina, which are useful for diagnosing and monitoring various retinal diseases. However, manual segmentation of OCTA images is a time-consuming and labor-intensive task, which motivates the development of automated segmentation methods. In this paper, we propose a novel multi-task learning method for OCTA segmentation, called OCTA-MTL, that leverages an image-to-DT (Distance Transform) branch and an adaptive loss combination strategy. The image-to-DT branch predicts the distance from each vessel voxel to the vessel surface, which can provide useful shape prior and boundary information for the segmentation task. The adaptive loss combination strategy dynamically adjusts the loss weights according to the inverse of the average loss values of each task, to balance the learning process and avoid the dominance of one task over the other. We evaluate our method on the ROSE-2 dataset its superiority in terms of segmentation performance against two baseline methods: a single-task segmentation method and a multi-task segmentation method with a fixed loss combination.
\vspace{-3mm}
\end{abstract}

\section{Introduction}
In recent years, Optical Coherence Tomography (OCT) has emerged as a powerful non-invasive imaging technique that has revolutionized the field of ophthalmology by providing high-resolution cross-sectional images of biological tissues, particularly the retina \cite{decarlo2015review,feucht2019optical}. One of the pivotal downstream analysis of OCT angiography (OCTA) lies in segmentation, a process that involves extracting and delineating Henle's fiber layer (HFL) and retinal blood vessels within these images \cite{octadataset2020}. OCT segmentation plays a critical role in diagnosing and monitoring various retinal diseases, such as diabetic retinopathy and neovascularization, while aiding in understanding complex structures within the images \cite{automatic2021}. Unfortunately, manual segmentation of OCTA images by clinicians is a time-consuming and labor-intensive task. Consequently, this has sparked interest in the development of automated segmentation methods \cite{octadataset2020,automatic2021,integrated2022}. Deep learning, a cornerstone of modern medical image analysis, unsuprisingly have paved the way in automatic segmentation of OCTA images \cite{machine2021}. Despite their prowess, deep models are commonly trained via single objective functions such as binary cross entropy (BCE) loss or dice loss \cite{transunet,fourier2022,Moeskops_2016}. Such singular objective functions may not capture the rich complexity and diversity of representations that can arise from OCTA segmentation, especially for tubular structures such as vessels \cite{fourier2022}. Multi-task learning trains a single model to handle multiple related tasks together, leveraging shared knowledge to improve performance across all tasks \cite{multi_task2}. By jointly optimizing on multiple tasks, the model can learn common patterns and representations, leading to enhanced efficiency and understanding of the data \cite{multi_task1}. Previous studies have shown that deep multi-task methods can significantly improve performance and generalization of medical image analysis models \cite{multi_task1}.

In this study, we introduce a multi-task approach (OCTA-MTL) that effectively captures complex and diverse features of OCTA images. Unlike vanilla segmentation architectures, OCTA-MTL leverages an image-to-DT (Distance Transform) branch that computes the distance from each vessel voxel to the vessel surface which can provide useful shape prior and boundary information for the encoder during training. Image-to-DT branch and segmentation head employs a joint encoder, and the aggregate network is simultaneously optimized via pixel-wise (regression) and binary cross-entropy (classification) losses. We then introduce a novel adaptive combination strategy that controls relative weighings of these loss functions. Demonstrations on ROSE-2 dataset \cite{octadataset2020} shows OCTA-MTL's superiority in terms of accuracy, robustness, and generalization against a single-task (Unet) method and a multi-task variant that employs an existing loss weighting strategy.

\begin{figure}[!t]
\vspace{-2.5ex}
\centerline{\includegraphics[width=0.9\textwidth]{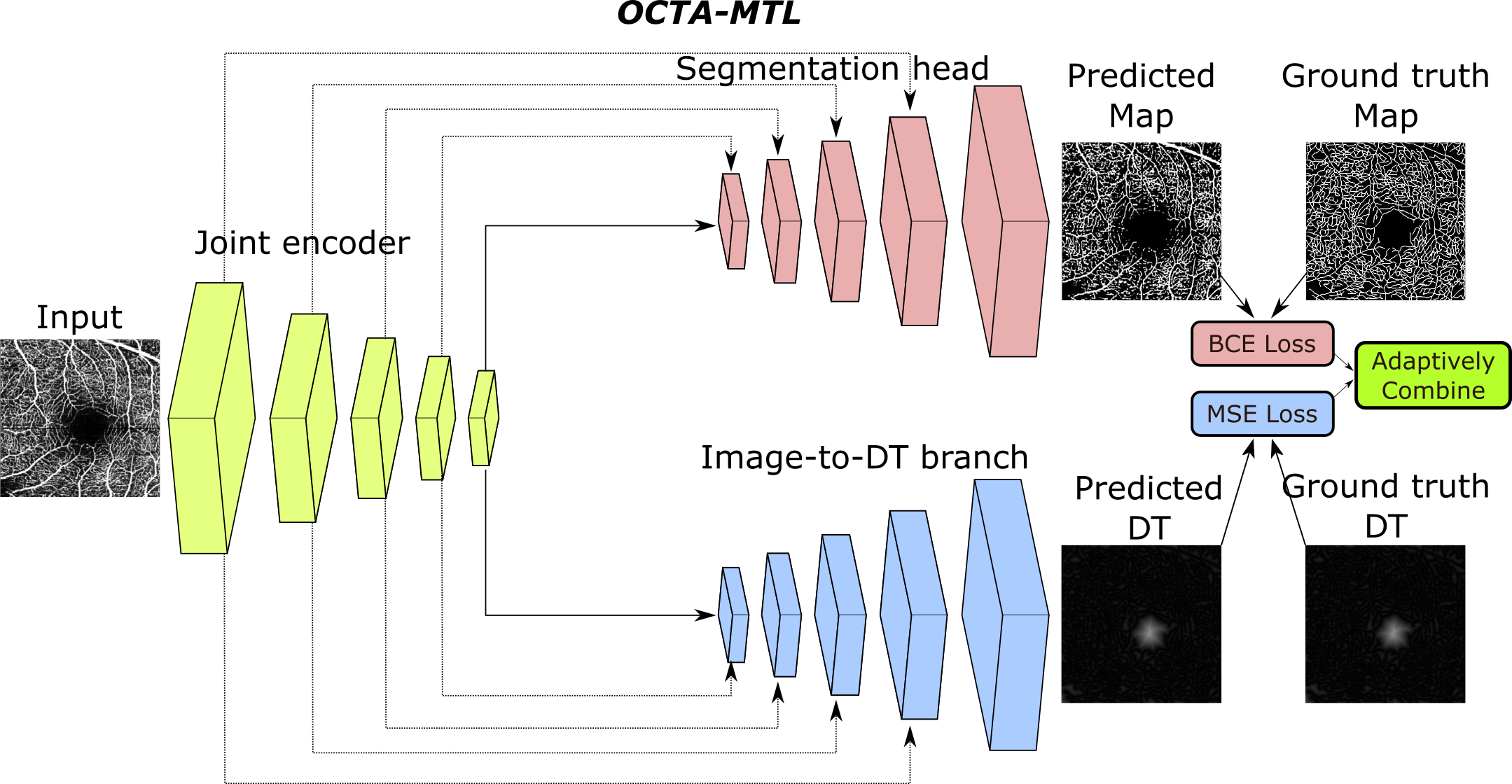}}
 \caption{The overall architecture of OCTA-MTL, a novel multi-task learning method for OCTA segmentation.}
 \label{fig:main}

\vspace{-3.5ex}
\end{figure}

\section{Methods}
\subsection{OCTA-MTL}
 OCTA-MTL, consists of two main components: a segmentation head and an image-to-DT branch. The segmentation head is responsible for predicting the binary mask of the vessels, whereas the image-to-DT branch is responsible for predicting the distance transform (DT) of the vessels. DT is a mathematical operator that assigns to each voxel in the image the distance to the nearest vessel surface.  DT can provide useful shape prior and boundary information for the segmentation task, as well as facilitate the evaluation of segmentation quality. Both the segmentation head and the image-to-DT branch share a common encoder, which extracts high-level features from the input OCTA image. The overall architecture of OCTA-MTL is illustrated in Figure \ref{fig:main}.

% \textbf{Encoder} The encoder of OCTA-MTL is based on the residual U-Net \cite{resu_net}, a popular encoder-decoder architecture widely adopted for medical image segmentation. The output of the encoder is a feature map of size 64x64x512, which is fed into both the segmentation head and the image-to-DT branch.

\textbf{Segmentation head} The segmentation head of OCTA-MTL is a decoder that reconstructs the binary mask of the vessels from the encoder output. The final output of the decoder is passed through a sigmoid activation function to obtain the probability of each voxel belonging to the vessel class. Here, a binary cross-entropy (BCE) loss is calculated.

\textbf{Image-to-DT branch} The image-to-DT branch of OCTA-MTL is a dense-pixel regression network that predicts the DT of the vessels from the encoder output. The predicted DT is then compared with the ground truth DT, which is computed from the ground truth binary mask of the vessels using the Euclidean distance transform algorithm \cite{borgefors1986distance}.

\textbf{Adaptive loss combination} To train OCTA-MTL, we use a combination of two loss functions: a binary cross-entropy (BCE) loss for the segmentation task and a mean squared error (MSE) loss for the image-to-DT task. Simply adding these two loss functions may not result in an optimal performance, as the scale and importance of each task may vary \cite{competing}. Therefore, we introduce a novel adaptive loss combination strategy that controls the relative weighting of these loss functions. Our strategy is based on the idea of dynamically adjusting the loss weights according to the inverse of the average loss values for each task. Specifically, we define a loss weight as follows:

\begin{align}
  \alpha = \mathbb{E}[||\mathcal{L}_{BCE}||] /\mathbb{E}[ ||\mathcal{L}_{MSE}||]
\end{align}

where $||\mathcal{L}_{BCE}||$ and $||\mathcal{L}_{MSE}||$ are the loss values of BCE and MSE. The intuition behind this strategy is that the task with a higher average loss value should be assigned a lower weight, as it indicates that the task is more difficult or less important than the other task. By doing so, we can balance the learning process of both tasks and avoid the dominance of one task over the other. In implementation, value of $\alpha$ is empirically estimated and updated within a batch. The final loss function for OCTA-MTL is then calculated via weighing MSE loss with $\alpha$ and linearly combining it with BCE loss:
\begin{align}
\mathcal{L} = ||L_{BCE}|| + \alpha \cdot ||L_{BCE}||
\end{align}

\section{Results}
\begin{figure}[!t]
\vspace{-1.5ex}
\centerline{\includegraphics[width=0.9\textwidth]{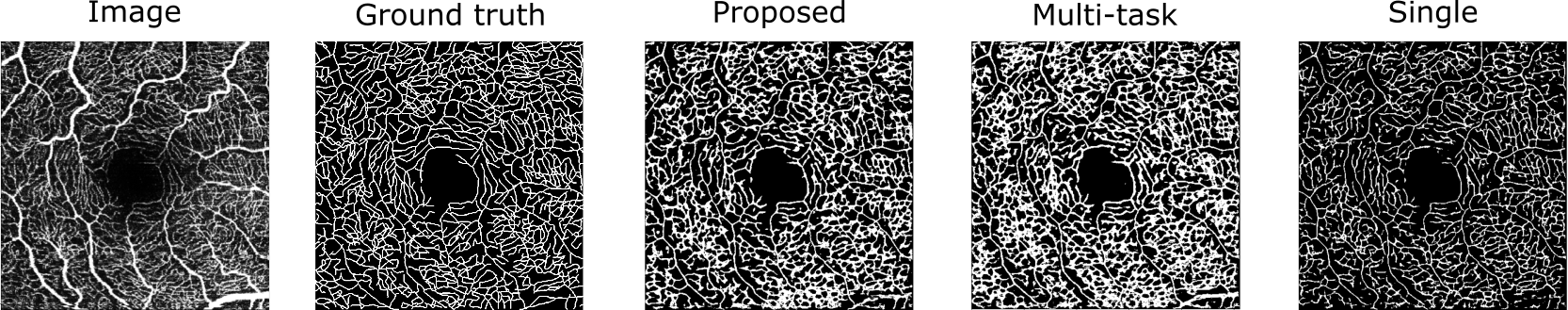}}
 \caption{OCTA images, ground truth segmentation maps, and maps predicted by the competing methods are shown in the figure.}
 \label{fig:result}
\end{figure}

We evaluated our proposed method, OCTA-MTL, on the ROSE-2 dataset \cite{octadataset2020}, and compared it with two baseline methods: a single-task segmentation method (Single Unet) \cite{unet2015} and a multi-task segmentation method that combined losses via an other common algorithm (Multi-task) \cite{competing}. We used the dice coefficient (Dice) and the intersection over union (IoU) scores as the evaluation metrics. Table \ref{tab:table1} shows that our method achieved the highest Dice and IoU scores, indicating its accuracy and consistency. Figure \ref{fig:result} shows a qualitative comparison of the segmentation results for a sample test slice. The proposed method produced more accurate and consistent segmentation results than the other methods, as it can better capture the complex and diverse features of the vessels, while avoiding the false positives and false negatives that are present in the other methods, especially in the regions with low contrast or high noise. These findings indicate that the proposed method benefits from effective leverage of image-to-DT branch and adaptive loss combination strategy.

\begin{table*}[]
\centering
\caption{Quantitative comparison of the segmentation performance in terms of average dice coefficient (Dice) and the intersection over union (IoU) scores of competing methods on test set.}
    \label{tab:table1}
\begin{tabular}{ll|l|l}
                          & Proposed & Multi-task \cite{competing} & Single Unet \cite{unet2015} \\ \hline
\multicolumn{1}{l|}{Dice} & \textbf{0.57}    & 0.57       & 0.54   \\ \hline
\multicolumn{1}{l|}{IoU}  & \textbf{0.35}     & 0.30       & 0.20  
\end{tabular}
\end{table*}

\section{Discussion}

 In this paper, we proposed OCTA-MTL, a novel multi-task learning method for OCTA segmentation that leverages an image-to-DT branch and an adaptive loss combination strategy. We showed that our method can effectively capture the complex and diverse features of the OCTA images, and improve the segmentation performance and generalization over the baseline methods. The proposed image-to-DT branch provides useful shape prior and boundary information for the segmentation task, while the adaptive loss combination strategy balances the learning process of both tasks and avoids the dominance of one task over the other. Our method has several limitations and directions for future work, such as evaluating it on other OCTA datasets, exploring transformer-based architectures and loss functions, and extending it to segment other structures in the OCTA images.

\section{Potential Negative Societal Impact}
This study aims to utilize a multi-task learning approach for improved performance and reliability in medical image segmentation. Consequently, our work does not put society in a negative position. This work could be utilized to improve the effectiveness of automated image analysis tasks in the future. Lastly, we are confident that the suggested approach does not exploit any potential bias in the data.

\bibliography{refs}

%%%%%%%%%%%%%%%%%%%%%%%%%%%%%%%%%%%%%%%%%%%%%%%%%%%%%%%%%%%%

\end{document}